\documentclass[sigconf]{acmart}

\copyrightyear{2023}
\acmYear{2023}
\setcopyright{acmlicensed}
\acmConference[MM '23]{Proceedings of the 31st ACM International Conference on Multimedia}{October 29-November 3, 2023}{Ottawa, ON, Canada}
\acmBooktitle{Proceedings of the 31st ACM International Conference on Multimedia (MM '23), October 29-November 3, 2023, Ottawa, ON, Canada}
\acmPrice{15.00}
\acmDOI{10.1145/3581783.3612474}
\acmISBN{979-8-4007-0108-5/23/10}
\usepackage[T1]{fontenc}
\usepackage{aecompl}

\usepackage{amsmath, amsthm, amssymb}
\usepackage{bbm}
\usepackage{bbding}
\usepackage{threeparttable}
\usepackage{subfigure}
\usepackage{makecell}
\usepackage{multirow}
\usepackage{graphicx}
\usepackage{float}
\usepackage{enumitem} 
\usepackage{enumerate}
\usepackage{pifont}
\usepackage{caption}
\usepackage{balance}
\usepackage{setspace}
\usepackage{graphicx}
\usepackage[T1]{fontenc}
\usepackage{color}

\usepackage{aecompl}

\usepackage{xcolor}
\usepackage{hyperref}
\hypersetup{colorlinks,
      linkcolor=ACMPurple,
      citecolor=ACMPurple,
      urlcolor=ACMDarkBlue,
   filecolor=ACMDarkBlue
}
\usepackage{stmaryrd}
\usepackage{enumitem}
\usepackage{colortbl}

\usepackage{algorithm}
\usepackage{booktabs}
\usepackage{makecell}
\usepackage{multirow}
\usepackage{array}
\usepackage{booktabs}
\usepackage{algorithmic}
\usepackage{cleveref}

\DeclareMathOperator*{\argmin}{arg\,min}
\newcommand{\black}[1]{\textcolor{black}{#1}}

\def\ie{\textit{i.e.}}

\def\eg{\textit{e.g.}}
\def\etal{\textit{et al.}}

\begin{document}


\title{A Four-Pronged Defense Against Byzantine Attacks in Federated Learning}

\author{Wei Wan}
\email{wanwei_0303@hust.edu.cn}
\affiliation{%
  \institution{School of Cyber Science and Engineering, Huazhong University of Science and Technology} 
  \country{}
}

\authornote{National Engineering Research Center for Big Data Technology and System}
\authornote{Services Computing Technology and System Lab}
\authornote{Hubei Key Laboratory of Distributed System Security}
\authornote{Hubei Engineering Research Center on Big Data Security}

\author{Shengshan Hu}
\email{hushengshan@hust.edu.cn}
\affiliation{%
  \institution{School of Cyber Science and Engineering, Huazhong University of Science and Technology}
  \city{}
  \country{}
}
\authornotemark[1]
\authornotemark[2]
\authornotemark[3]
\authornotemark[4]

\author{Minghui Li}
\email{minghuili@hust.edu.cn}
\affiliation{%
  \institution{School of Software Engineering, Huazhong University of Science and Technology}
  \city{}
  \country{}
}

\author{Jianrong Lu}
\email{lujianrong@hust.edu.cn}
\affiliation{%
  \institution{School of Cyber Science and Engineering, Huazhong University of Science and Technology}
  \city{}
  \country{}
}
\authornotemark[1]
\authornotemark[2]
\authornotemark[3]
\authornotemark[4]

\author{Longling Zhang}
\email{longlingzhang@hust.edu.cn}
\affiliation{%
  \institution{School of Cyber Science and Engineering, Huazhong University of Science and Technology}
  \city{}
  \country{}
}
\authornotemark[1]
\authornotemark[2]
\authornotemark[3]
\authornotemark[4]

\author{Leo Yu Zhang}
\email{leo.zhang@griffith.edu.au}
\affiliation{%
  \institution{School of Information and Communication Technology, Griffith University}
  \city{}
  \country{}
}

\author{Hai Jin}
\email{hjin@hust.edu.cn}
\affiliation{%
  \institution{School of Computer Science and Technology, Huazhong University of Science and Technology}
  \city{}
  \country{}
}
\authornotemark[1]
\authornotemark[2]
\authornote{Cluster and Grid Computing Lab}

\renewcommand{\shortauthors}{Wei Wan et al.}

\begin{abstract}

\textit{Federated learning} (FL) is a nascent distributed learning paradigm to train a shared global model without violating users' privacy.   FL has been shown to be vulnerable to various Byzantine attacks, where malicious participants could independently or collusively upload well-crafted updates to deteriorate the performance of the global model. However, existing defenses could only mitigate part of Byzantine attacks,  without providing an all-sided shield for FL. It is difficult to simply combine them as they rely on totally contradictory assumptions. 

In this paper, we propose FPD, a \underline{\textbf{f}}our-\underline{\textbf{p}}ronged \underline{\textbf{d}}efense against both non-colluding and colluding Byzantine attacks. Our main idea is to utilize absolute similarity to filter updates rather than relative similarity used in existingI works. To this end, we first propose a reliable client selection strategy to prevent the majority of threats in the bud. Then we design a simple but effective score-based detection method to mitigate colluding attacks. Third, we construct an enhanced spectral-based outlier detector to accurately discard abnormal updates when the training data is \textit{not independent and identically distributed} (non-IID). Finally, we design update denoising to rectify the direction of the slightly noisy but harmful updates. The four sequentially combined modules can effectively reconcile the contradiction in addressing non-colluding and colluding Byzantine attacks. Extensive experiments over three benchmark image classification datasets against four state-of-the-art Byzantine attacks demonstrate that FPD drastically outperforms existing defenses in IID and non-IID scenarios (with $30\%$ improvement on model accuracy).

\end{abstract}




\maketitle
\section{Introduction}

\textit{Federated learning} (FL)~\cite{FedAvg, LongtailFL} emerges as a new distributed machine learning paradigm recently, where the training data and the learning process are fully controlled by the clients, thus alleviating the privacy concern. 
However, due to its decentralized nature, FL is found to be highly vulnerable to Byzantine attacks, where malicious participants contribute poisoned updates to damage the global model. Generally, Byzantine attacks can be categorized into non-colluding attacks~\cite{PCA,Zeno} where attackers upload malicious updates independently, and colluding attacks~\cite{ALittleIsEnough,IPM,LocalModelPoisoning,AGRT,MPAF,IJCAI-ZHT} where attackers share information (\eg, training data, and model updates) to each other and collusively design well-crafted updates. In particular, colluding attackers tend to upload  similar or totally identical updates to avoid being treated as outliers~\cite{SparseFed}.


To defend against these two kinds of attacks, massive defensive schemes have been proposed in recent years. For the non-colluding attacks, existing defenses, such as Krum~\cite{Krum}, FABA~\cite{FABA}, Median~\cite{TrimmedMean}, FedInv~\cite{FedInv}, AFA~\cite{AFA}, manage to remove or circumvent the outliers based on the intuition that benign updates are much similar to each other due to the same optimization objective, while the malicious ones can be considered as outliers. To resist colluding attacks, existing works like \black{FoolsGold~\cite{Sybils}, LOF~\cite{WeightAttack}, and Contra}~\cite{Contra} propose to punish the relatively similar updates by distributing smaller weights in the aggregation stage. Unfortunately, these defenses (or simply combining them) cannot mitigate non-colluding and colluding attacks simultaneously, \black{since the intuitions behind them are almost opposite arguing whether the malicious updates are similar to each other.}

Recent studies like \black{LFR}~\cite{LocalModelPoisoning}, Zeno~\cite{Zeno}, FLTrust~\cite{FLTrust}, DiverseFL~\cite{DiverseFL} attempt to defend against both attacks simultaneously. Instead of relying on the distribution of the updates, they turn to an auxiliary dataset to validate the performance (\eg, loss or accuracy) of each update~\cite{LocalModelPoisoning,Zeno}, or construct a reliable update as a reference~\cite{FLTrust,DiverseFL}. These performance-based defenses hold that malicious updates inevitably degrade model performance in a degree. Although performing much better in non-colluding and part of colluding scenarios, these defenses fail to work when malicious updates are slightly noised but harmful (\eg, LIE attack~\cite{ALittleIsEnough}), especially when the data is \textit{not independent and identically distributed} (non-IID). Moreover, \black{the assumption of possessing an auxiliary dataset will violate users' privacy as they usually require that the auxiliary dataset has the same distribution as the clients' local training datasets.} In summary, an effective defense providing an all-sided shield for FL is still missing yet.

To tackle these issues, we propose FPD, a \underline{\textbf{f}}our-\underline{\textbf{p}}ronged \underline{\textbf{d}}efense against both non-colluding and colluding Byzantine attacks. Our main observation is that \black{the contradictory intuitions behind the existing two kinds of schemes arise because both of them rely on the relative similarity between updates due to the lack of a gold standard to evaluate each update in FL. In light of this, we propose to construct an artificial gold standard, which is an empirically determined threshold, to form absolute similarity that can be used to detect colluding attacks. Meanwhile, non-colluding attacks can still be detected based on relative similarity. In this way, the contradictory of solely exploiting relative similarity can be reconciled naturally. }
Specifically, we propose two defense modules relying on absolute similarity and relative similarity to defend against colluding attacks and non-colluding attacks, respectively. In addition, we design a reliable client selection strategy to prevent the majority of threats and the update denoising method to rectify the update directions, in order to further \black{alleviate the impact of colluding attacks}.

In summary, we offer the following contributions:

\begin{itemize}
    \item We propose a new FL defense scheme FPD, which is effective in defending against  non-colluding and colluding Byzantine attacks simultaneously.
    
    \item We propose two novel auxiliary defense modules (\ie, reliable client selection and update denoising) to further enhance the defense ability.
    
    

    \item We demonstrate the advantage of FPD via extensive experiments on three benchmark datasets against four state-of-the-art attacks. Compared with five distinguished defenses, our scheme achieves the best performance in both IID and non-IID scenarios.
    
\end{itemize}

\section{Background}
\subsection{Federated Learning}


We consider a general FL system, consisting of a central server and $K$ clients. Each client $k\in [K]$ has a dataset $D_{k}$, the size of which is denoted as  $|D_{k}|=n_{k}$. It is worth noting that each local dataset may be subject to a different distribution, that is, the clients' data may be distributed in a non-IID  way. The clients aim to collaboratively train a shared global model $\boldsymbol{w}$. Apparently, the problem can be solved via minimizing the empirical loss, \ie, $\argmin_{\boldsymbol{w}} f(D,\boldsymbol{w})$, where $D=\bigcup_{k=1}^{K} D_{k}$ and $f(D,\boldsymbol{w})$ is a loss function (\eg, mean absolute error, cross‐entropy). However, the optimization requires all the clients to share their raw data to a central server, which would result in a serious threat to client's privacy. Instead, FL obtains $\boldsymbol{w}$ by optimizing $\argmin_{\boldsymbol{w}} \sum_{k=1}^k f(D_{k},\boldsymbol{w})$. Specifically, the FL system iteratively performs the following three steps until the global model converges:
\begin{itemize}
	\item \textbf{Step I:} In the $t$-th iteration, the central server broadcasts a global model $\boldsymbol{w_{t}}$ to the clients;
	\item \textbf{Step II:} After receiving $\boldsymbol{w_{t}}$, each client $k$ trains a new local model $\boldsymbol{w_{t}^{k}}$ over $D_{k}$ by solving the optimization problem $\argmin_{\boldsymbol{w_{t}^k}} f(D_{k},\boldsymbol{w_{t}^k})$ and then uploads the local model update $\boldsymbol{g_{t}^k} := \boldsymbol{w_{t}^k} - \boldsymbol{w_{t}}$ to the server;
	\item \textbf{Step III:} The server aggregates all the local updates according to client's proportional dataset size as follow:
	\begin{equation}
	\boldsymbol{w_{t+1}}\gets \boldsymbol{w_{t}}+\sum_{k=1}^{K} \frac{n_{k}}{n} \boldsymbol{g_{t}^{k}}, \text{where} \ n=\sum_{k=1}^{K}n_{k}.
	\end{equation}
\end{itemize}

\section{Threat Model}
\subsection{Attack Model}
Following previous studies~\cite{FedInv,Sybils,ALittleIsEnough}, we consider a strong attack model where an adversary controls $f$ out of the total $K$ participants. The adversary can arbitrarily manipulate the data and the updates of the controlled clients. The goal of the adversary is to upload well-crafted malicious updates via the controlled clients to damage the global model accuracy. The controlled clients can collude with each other, and the adversary may possess the knowledge (\eg, the local updates) of other uncontrolled clients so as to initiate stronger attacks.
\subsection{Defense Model}
To design a practical defense, we cast away the following unrealistic assumptions that existing defenses rely on.

\begin{itemize}
	\item \textbf{Training dataset sizes.} Recently proposed defense~\cite{FedInv} assumes that the training dataset sizes of all the clients are known by the central server so that a fair weight distribution mechanism can be built. However, clients can arbitrarily report the sizes due to the distributed nature~\cite{WeightAttack,FLSurvey}.
	
	\item \textbf{Number of attackers.} Many defenses~\cite{Krum,FABA,AGRT,FedInv} assume that the central server knows the number of attackers so as to determine how many updates should be removed. Nevertheless, the clients in FL are dynamically changing and cannot be determined in advance.
	
	\item \textbf{Auxiliary dataset.} Many defenses~\cite{FLTrust,LocalModelPoisoning,Zeno,DiverseFL,RobustFL} rely on an auxiliary dataset whose distribution is the same as that of the clients, to evaluate the performance of the local updates. However, this assumption undoubtedly violates users' privacy. 
	
\end{itemize}

On the contrary, our defense makes minimum assumptions. The only information the central server holds is the local updates uploaded by the clients. The goal of our defense is to achieve the competitive model accuracy in both non-colluding and colluding scenarios.

\begin{figure*}[t]
\centerline{\includegraphics[width=1.96\columnwidth]{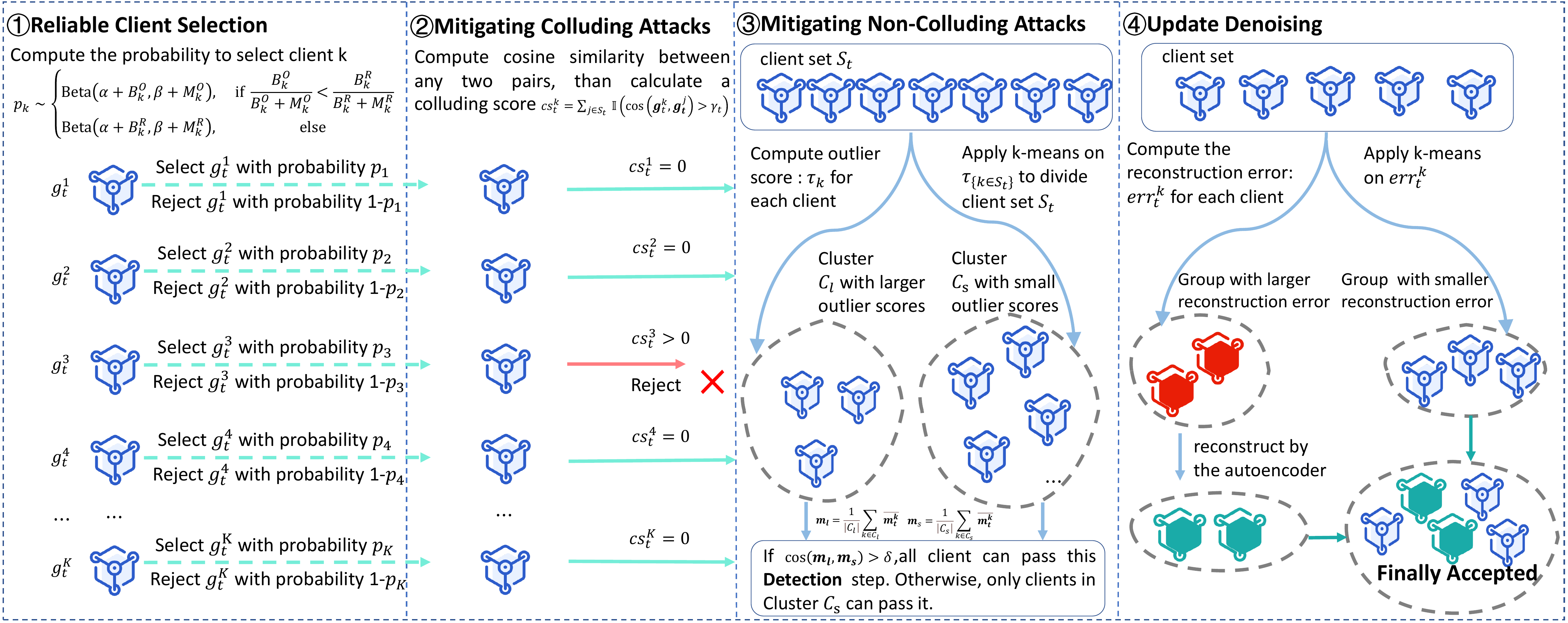}}
\caption{The workflow of our proposed FPD}
\label{fig:OurScheme}
\end{figure*}

\section{FPD: a Four-Pronged Defense against Byzantine Attacks}
\subsection{Motivation and Overview of FPD}
After reviewing state-of-the-art defenses, we find that none of them can fully protect FL. Specifically, the colluding oriented defenses cannot defend against non-colluding attacks, and vice versa.  Simply combining these two kinds of defenses seems promising, but they rely on totally contradictory assumptions. \black{The former assumes that malicious updates are relatively similar, while the latter considers benign updates are more compact.} Since all of these defenses employ relative similarity as a metric to filter out outliers, a combination of them inevitably leads to the rejection of benign updates in either colluding scenario or non-colluding scenario. 
Although the performance based defenses try to handle both of these two attacks, they are unable to detect malicious updates which are slightly perturbed but maintain  toxicity. 

To reconcile such a dilemma, we propose using absolute similarity to filter out extremely similar updates, and then employ an outlier detector based on relative similarity to discard abnormal updates. Furthermore, we propose two auxiliary defense modules (\ie, the client selection and the update denoising) \black{to further restrain the attack space of the poisoned updates, thus making it easier to filter out colluding and non-colluding poisoned updates}.
As shown in Fig.~\ref{fig:OurScheme}, our proposed FPD consists of the following four steps.

\begin{itemize}
	\item \textbf{Step I: Reliable Client Selection.} Instead of randomly selecting a subset of clients to participate in each iteration, the central server  selects the reliable clients who are more likely to contribute high quality updates according to the historical performance of each participant.
	
	\item \textbf{Step II: Mitigating Colluding Attacks.} The central server detects and rejects the updates that are excessively similar in the direction space once receiving all the local updates from the currently selected clients.
	
	\item \textbf{Step III: Mitigating Non-Colluding Attacks.} The central server detects and rejects the outliers via a spectral-based outlier detector.
	
	\item \textbf{Step IV: Update Denoising.} The central server applies an autoencoder to reconstruct the malicious updates that are too similar to benign ones to detect.
	
\end{itemize}

\newtheorem{remark}{Remark}
\begin{remark}
Step I ensures that most of the malicious clients cannot participant in FL at all, in other words, only a limited number of compromised clients have a chance to poison the global model. Step II prohibits the adversary from designing excessively similar malicious updates, enhancing the difficulty of launching a covert attack. Step III guarantees that any update far from the overall distribution would be discarded. Step IV is designed to rectify the direction of the slightly noised but harmful updates. Note that   Steps I and IV are directly dependent on the detection capacity of the Steps II and III, which 
inform the server whether an update is benign or malicious.
\end{remark}

\subsection{Reliable Client Selection}
Client selection is widely studied in the FL community, through which the researchers aim to reduce the communication overhead~\cite{ClientSelectionCommunication}, solve the data heterogeneity challenge~\cite{ClientSelectionNonIID}, and deal with the resource constrained FL scenarios~\cite{ClientSelectionResource}. However, it is rarely considered in the Byzantine-robust FL field. To the best of our knowledge, there are only two related defenses. In AFA~\cite{AFA}, the authors propose a blocking mechanism to forbid the clients to participate in the subsequent iterations once they have shared sufficient bad updates. Recently, Wan~\etal~\cite{MABRFL} proposed MAB-RFL, which applies a Beta distribution to estimate the probability of each client providing a benign update in the current iteration. However, both the defenses only focus on the overall performance of each client without taking their recent behaviors into account. Therefore, the attackers, in the early stages, can pretend to be benign clients by uploading well-trained updates to earn trust from the central server, and thus they will be constantly selected even though their latest updates are malicious.

Based on the above observation, we propose a new client selection strategy which considers both the overall and the recent performance of each client such that:

(i) The client who has uploaded too many malicious updates is selected with a low probability even though it has performed well in the recent iterations; 

(ii) The client who has contributed substantial benign updates while performing badly in the recent iterations is also selected with a low probability; 

(iii) Only the client who persistently shares benign updates is selected with a high probability.

Formally, in the client selection stage, the central server selects each client $k$ with the probability:
\begin{equation}\label{eq:sampling}
p_{k}\sim\left\{\begin{matrix}
 \text{Beta}(\alpha+B_{k}^{O},\beta+M_{k}^{O}), & \text{if~} \frac{B_{k}^{O}}{B_{k}^{O}+M_{k}^{O}}< \frac{B_{k}^{R}}{B_{k}^{R}+M_{k}^{R}}  \\
 \text{Beta}(\alpha+B_{k}^{R},\beta+M_{k}^{R}), &\text{else}
\end{matrix}\right.
\end{equation}
where $\alpha$ and $\beta$ are prior parameters. $B_{k}^{O}$ and $M_{k}^{O}$ denote the overall frequencies that the local updates from client $k$ are identified as benign and malicious respectively. Analogously, $B_{k}^{R}$ and $M_{k}^{R}$ indicate the recent frequencies. In this paper, we define the "recent" as the latest $10$ iterations a specific client is selected.

Note that the central server possesses limited information about a client's identity (\ie, benign or malicious) in the early iterations, thus it nearly makes a random choice, which deteriorates the convergence rate. As a remedy to this concern, we propose a bootstrap trick, by allowing all the clients to participate in the training in the first $10$ iterations so as to fully understand their identities.

\subsection{Mitigating Colluding Attacks}
Recently, colluding attacks have aroused extensive attention for its effectiveness in designing covert but powerful Byzantine attacks. For example, LIE attack~\cite{ALittleIsEnough} adds well-crafted noise, which is tiny enough to circumvent the defense while huge enough to degrade the global model accuracy, to a benign update. IPM attack~\cite{IPM} reverses the direction of a benign update in order to maximize the attack effect. Wan~\etal~\cite{WeightAttack} proposed free-riding attack, where attackers train local models on small amounts of data but declare large training set sizes so as to dominate the global model. Fang~\etal~\cite{LocalModelPoisoning}, and Shejwalkar~\etal~\cite{AGRT} proposed optimization based attacks respectively. Albeit different in implementation, all the attacks are based on a core idea, that is, the attackers should collude with each other to make the malicious updates as similar as possible or even totally identical. Colluding attack indeed poses a great threat to existing defenses as  verified by our experiments. The difficulty in defending against colluding attack lies in the following facts:

(i) Benign updates are inevitably got punished~\cite{Sybils,Contra,WeightAttack};

(ii) It is hard to reconcile colluding attack and non-colluding attack. 

To address these two challenges, we first propose a simple yet effective solution to mitigate colluding attacks by constructing absolute similarity. Specifically, we calculate a colluding score for each selected client $k$ as follow:
\begin{equation}\label{eq:collusionscore}
cs_{t}^{k} = \sum_{j\in S_{t}} \mathbb{I}(\cos(\boldsymbol{g_{t}^{k}},\boldsymbol{g_{t}^{j}})>\gamma_{t}),
\end{equation}
where $\mathbb{I}(\cdot)$ is the indicator function, $\cos(\cdot,\cdot)$ indicates the cosine similarity, $S_{t}$ is the selected client set in iteration $t$, $\gamma_{t}\in\left [-1, 1 \right ]$ denotes the tolerable cosine similarity threshold. As demonstrated in \cite{Sybils,Contra,MABRFL}, the benign updates will not be extremely similar to each other in the direction space even in an IID scenario, thus it is easy to set the threshold $\gamma_{t}$ (in our experiments we set $\gamma_{t}=0.8$) to filter out colluding attackers without affecting benign clients. Specifically, any client $k$ with a positive colluding score $cs_{t}^{k}$ will be regarded as malicious and rejected in this stage.

\subsection{Mitigating Non-Colluding Attacks}
In the scenario of non-colluding attacks, where malicious updates are quite different from each other in direction as well as magnitude, attackers can easily circumvent the detection of colluding attacks, which motivates the need of an additional abnormal detection step based on relative similarity. To this end, we borrow the idea from \cite{RobustMeanEstimation}, where a spectral-based outlier detector is proposed. At a high level, the algorithm first computes the top singular vector of a matrix composed of all the involved vectors. Then any vector whose projection onto the singular vector (\ie, the outlier score) is too large will be removed (by assuming the number of outliers is known in advance). Despite its good performance on several datasets with theoretical guarantee, it does not readily apply to our case due to the following challenges:
\begin{itemize}
	
	\item \textbf{Challenge I.} As demonstrated in the original paper, the method performs badly in non-IID scenario, which is the most representative feature in FL.
 
    \item \textbf{Challenge II.} The method is highly sensitive to the magnitudes of the involved vectors even in the IID scenario.
    

    \item \textbf{Challenge III.} The method requires the number of outliers. Unfortunately, FL is a dynamic distributed network where honest and malicious clients can join in and drop out arbitrarily.
\end{itemize}

To address Challenge I, we introduce momentum, which is shown to be effective to reduce the variance between updates~\cite{momentum,CC}.
In this way, an IID-like distribution can be built. 
Formally, we compute the momentum vector as: 
\begin{equation}
    \boldsymbol{m_{t}^{k}} = \boldsymbol{g_{t}^{k}} + \lambda^{t-t_{k}} \boldsymbol{m_{t_{k}}^{k}},
    \label{momentum}
\end{equation}
where $t_{k}$ is the latest selected iteration for client $k$, $\lambda\in(0,1)$ indicates the importance of historical information. Initially, we set $\boldsymbol{m_{t_{k}}^{k}}=\boldsymbol{0}$. Note that the iteration interval for a client being selected twice may be quite large, making the historical information that lies in the momentum vector $\boldsymbol{m_{t_{k}}^{k}}$ obsolete. Thus we multiply it by a smaller discount factor $\lambda^{t-t_{k}}$, rather than using $\lambda$ as existing works did.

To address Challenge II, we further normalize the momentum vector into an unit one:
\begin{equation}
    \boldsymbol{\overline{m_{t}^{k}}} =\frac{\boldsymbol{m_{t}}^{k}}{||\boldsymbol{m_{t}^{k}}||}.
    \label{normalize}
\end{equation}

In this way, the outlier-detector will focus on the direction only, without being affected by the magnitude.
Moreover, Eq.~(\ref{normalize}) also ensures that a single malicious update has a limited impact on the aggregation result, and a benign update with a small magnitude can contribute more information.

To address Challenge III, we apply the $k$-means algorithm to divide the normalized momentum vectors into two groups based on the outlier scores obtained by the outlier-detector \black{due to its simpleness and effectiveness}. Instead of simply treating the group with smaller outlier scores as being benign, we take the similarity between the two groups into consideration. Specifically, if the two groups are much similar (\ie, the cosine similarity exceeds a threshold $\delta$), it is very likely that all the updates are benign. In such a case, both groups will be kept for aggregation; otherwise, the group with larger outlier scores will be removed.

A detailed description for detecting non-colluding attack is summarized in Algorithm~\ref{alg:DDA}.

\begin{algorithm}[t]
\caption{Mitigating Non-Colluding Attacks}
\label{alg:DDA}
\textbf{Input}: Current iteration $t$, left clients $S_{t}$, latest selected iterations $\{t_{k}, k\in S_{t}\}$, local updates $\{\boldsymbol{g_{t}^{k}},k\in S_{t}\}$, momentum vectors $\{\boldsymbol{m_{t_{k}}^{k}}, k\in S_{t}\}$, acceptable difference between clusters $\delta$, importance of historical information $\lambda$\\
\textbf{Output}: Set of removed clients $R$
\begin{algorithmic}[1] 
\STATE Compute the normalized momentum vectors $\{\boldsymbol{\overline{m_{t}^{k}}},k\in S_{t}\}$ through Eq.~(\ref{momentum}) and Eq.~(\ref{normalize}).
\STATE Let $\boldsymbol{\mu} =\frac{1}{|S_{t}| } {\textstyle \sum_{k\in S_{t}}} \boldsymbol{\overline{m_{t}^{k}}}$.
\STATE Let $\boldsymbol{G}=[\boldsymbol{\overline{m_{t}^{k}}}-\boldsymbol{\mu} ]_{k\in S_{t}}$ be the matrix of centered vectors.
\STATE Let $\boldsymbol{v}$ be the top right singular vector of $\boldsymbol{G}$.
\STATE Compute \textit{outlier scores} defined as $\tau_{k}=((\boldsymbol{\overline{m_{t}^{k}}}-\boldsymbol{\mu})\cdot \boldsymbol{v})^{2}$.
\STATE Apply k-means on $\tau_{\{k\in S_{t}\}}$ to divide $S_{t}$ into two clusters $C_{l}$ with larger outlier scores and $C_{s}$ with smaller outlier scores.
\STATE Compute the mean vector of each cluster:\\
$\boldsymbol{m_{l}}=\frac{1}{|C_{l}|}  {\textstyle \sum_{k\in C_{l}}} \boldsymbol{\overline{m_{t}^{k}}}$;\\
$\boldsymbol{m_{s}}=\frac{1}{|C_{s}|}  {\textstyle \sum_{k\in C_{s}}} \boldsymbol{\overline{m_{t}^{k}}}.$
\IF {$cos(\boldsymbol{m_{l}},\boldsymbol{m_{s}})>\delta$}
\STATE Let the removed set $R=\varnothing$.
\ELSE
\STATE Let the removed set $R=C_{l}$.
\ENDIF
\RETURN $R$
\end{algorithmic}
\end{algorithm}

\subsection{Update Denoising}
Recent studies~\cite{ALittleIsEnough,AGRT,LocalModelPoisoning} show that attackers can upload well-crafted updates (by adding tiny noises to a benign update) that are extremely similar to benign ones to circumvent the defenses as well as maintain the attack effect. Distinguishing them from benign updates is really challenging. Therefore, instead of detecting and removing them, we denoise and utilize the slightly disturbed updates to facilitate the convergence. Specifically, we turn to an autoencoder to denoise the normalized momentum vectors that successfully get through the preceding detection steps, then the ones with large reconstruction errors will be reconstructed while the remaining vectors keep unchanged. Formally, the reconstruction error of client $k$ in iteration $t$ is given by:
\begin{equation}
    err_{t}^{k} =||\boldsymbol{\overline{m_{t}^{k}}}-ae(\boldsymbol{\overline{m_{t}^{k}}})||^{2},
    \label{reconstruction_error}
\end{equation}
where $ae(\cdot)$ represents the autoencoder. Then, we utilize the $k$-means algorithm to divide the normalized momentum vectors into two groups based on the reconstruction errors. The group with larger reconstruction errors will be denoised by the autoencoder, and the other group remains unchanged.

Note that training such an autoencoder does not require any raw data shared by participants, thus users' privacy is well protected. Instead, we use the historical reliable normalized momentum vectors (derived from local updates) as the training samples. Moreover, the dimension of the momentum vector $\boldsymbol{\overline{m_{t}^{k}}}$ (the same with that of the model weights) is generally quite large, making it time-consuming to train the autoencoder. Hence we only consider the weights between the last two layers, which are decisive for the classification results~\cite{VisualizingCNN}.

\begin{figure*}[t]
	\centering
	\subfigure[CIFAR-10]{\label{fig:lf_m}
		\includegraphics[width=0.6\columnwidth]{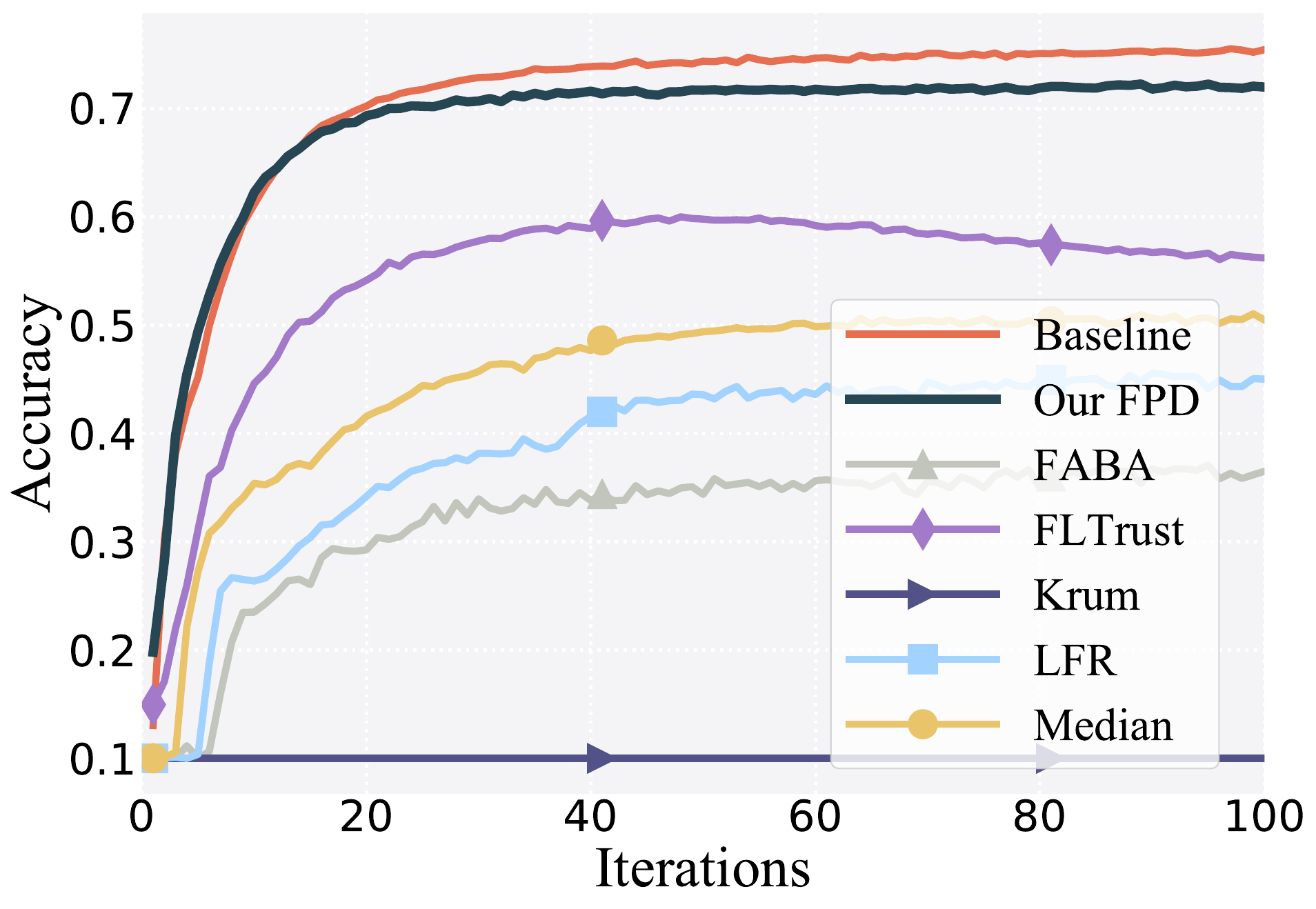}}\hspace{8mm}
	\subfigure[Fashion-MNIST]{\label{fig:lf_c}
		\includegraphics[width=0.6\columnwidth]{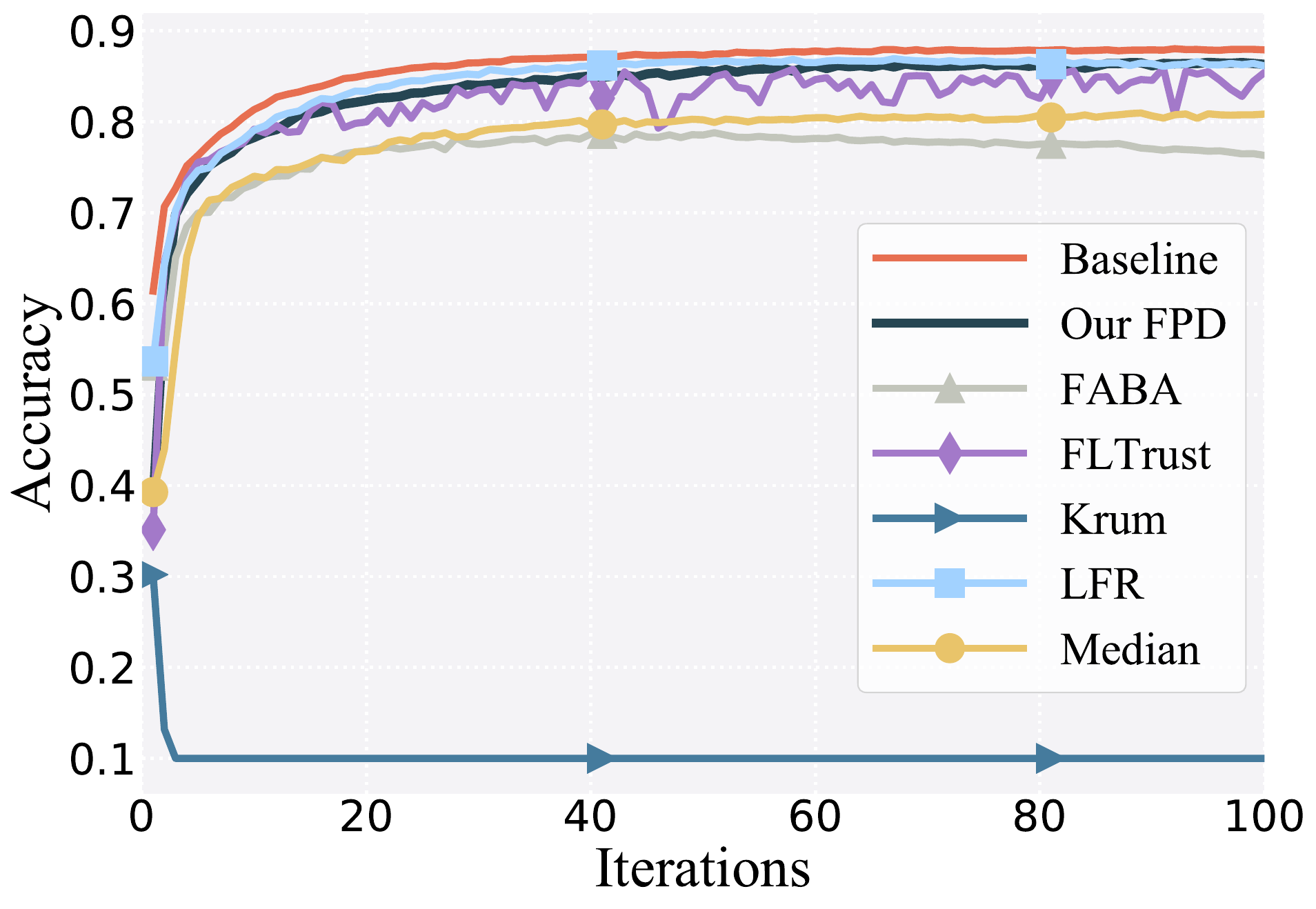}}\hspace{8mm}
	\subfigure[MNIST]{\label{fig:lf_c}
		\includegraphics[width=0.6\columnwidth]{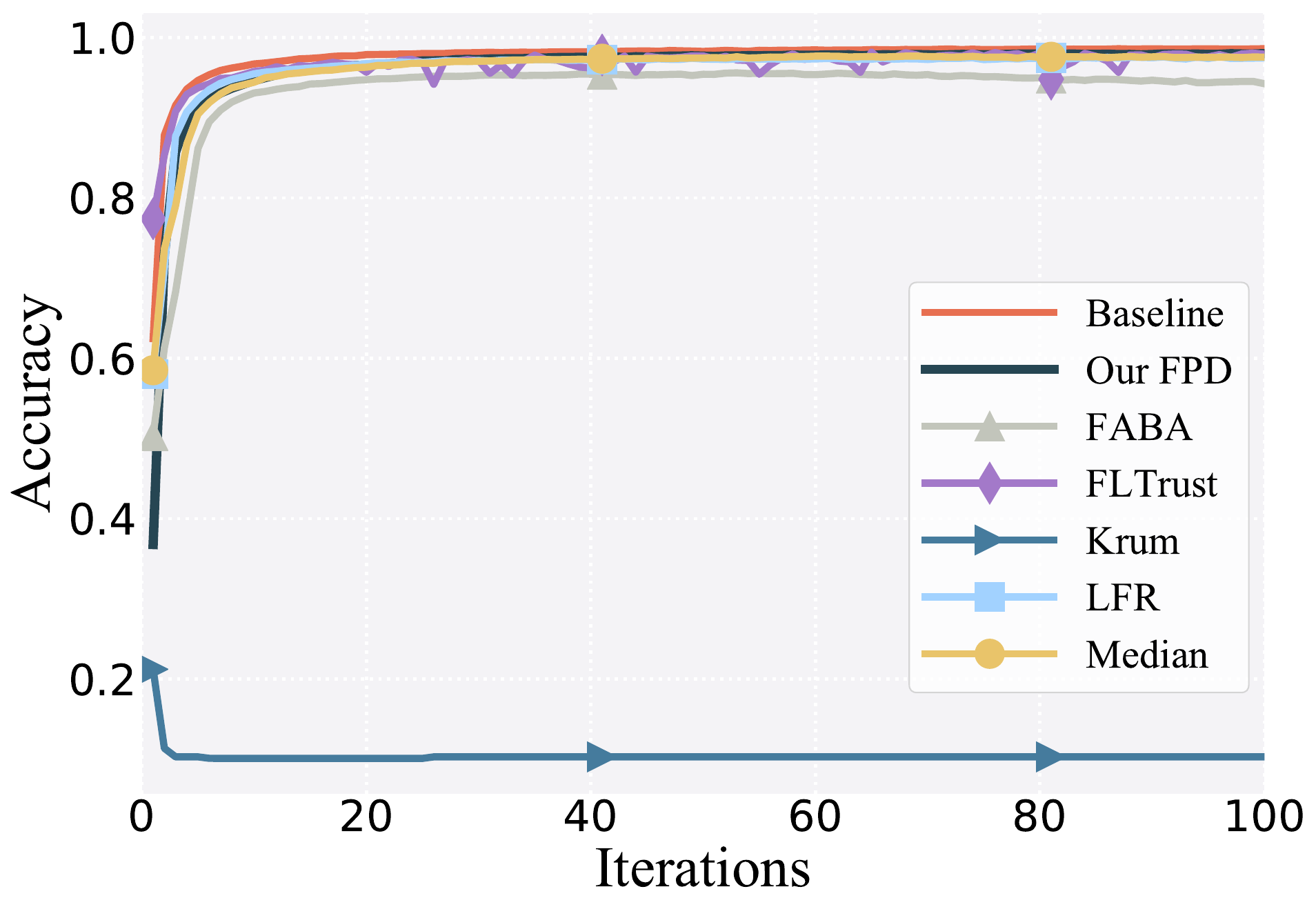}}
	\vspace{-2mm}
	\caption{Model accuracy under LIE attack}
	\label{fig:LIE}
\end{figure*}

\begin{figure*}[t]
	\centering
	\subfigure[CIFAR-10]{\label{fig:lf_m}
		\includegraphics[width=0.6\columnwidth]{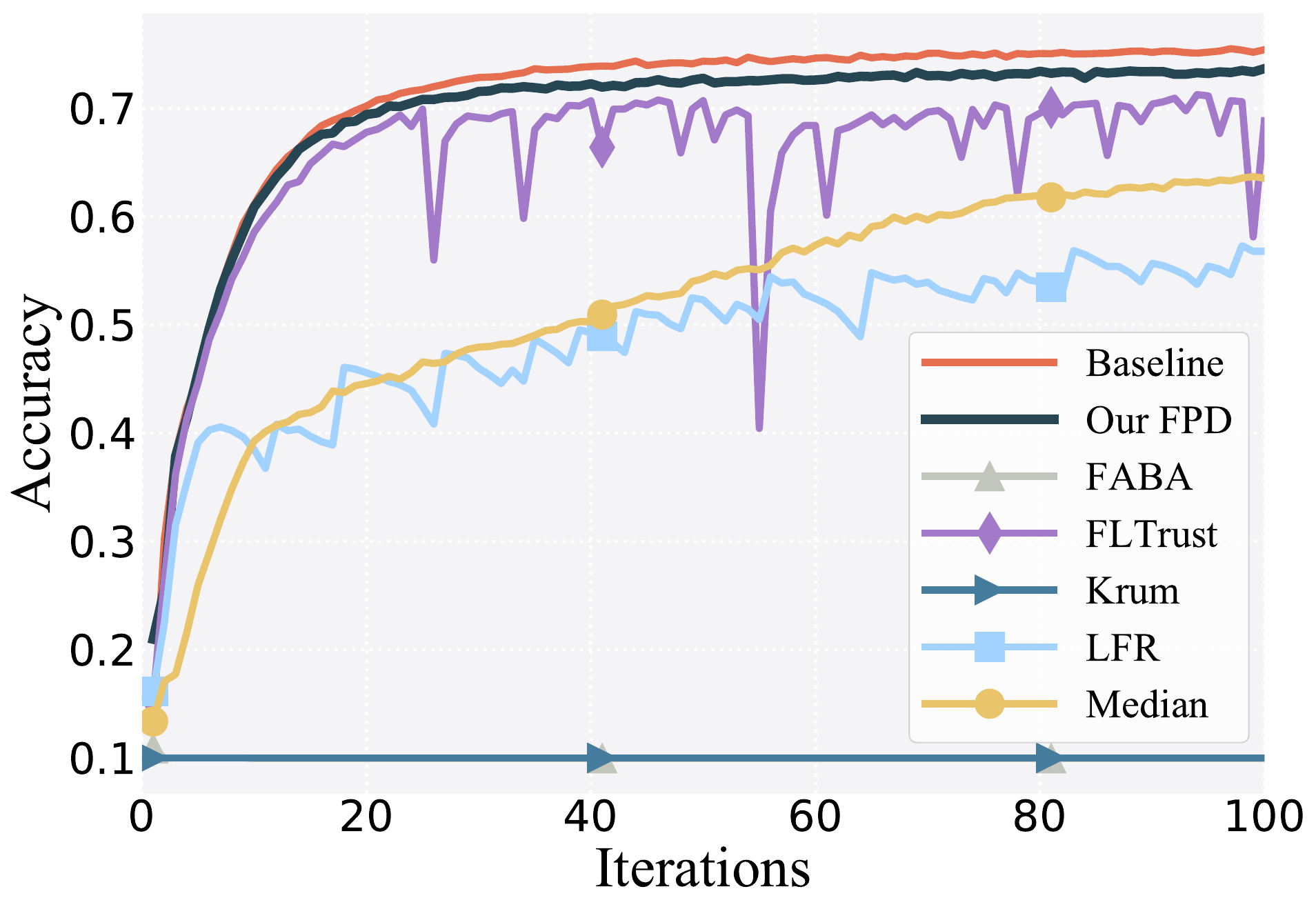}}\hspace{8mm}
	\subfigure[Fashion-MNIST]{\label{fig:lf_c}
		\includegraphics[width=0.6\columnwidth]{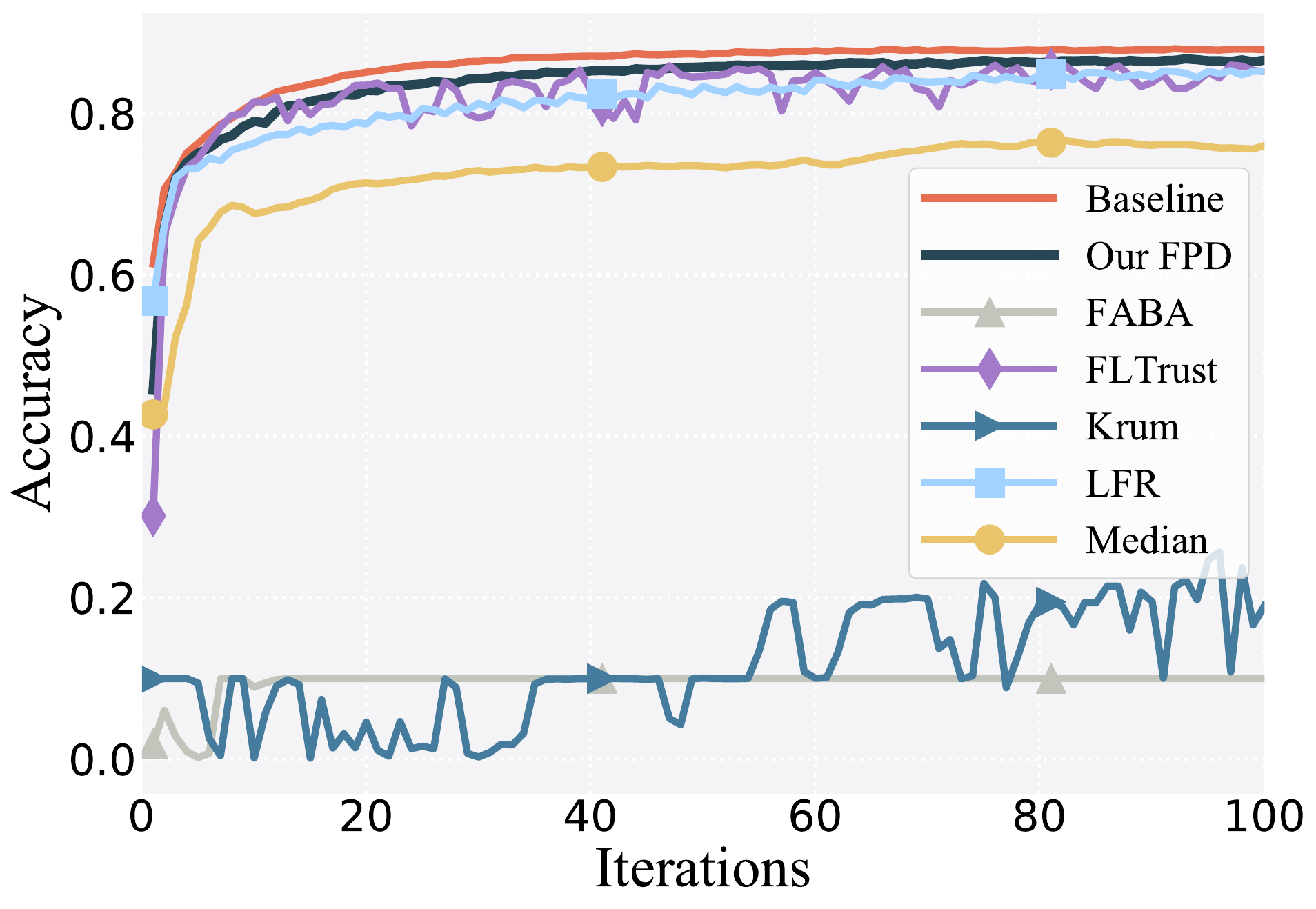}}\hspace{8mm}
	\subfigure[MNIST]{\label{fig:lf_c}
		\includegraphics[width=0.6\columnwidth]{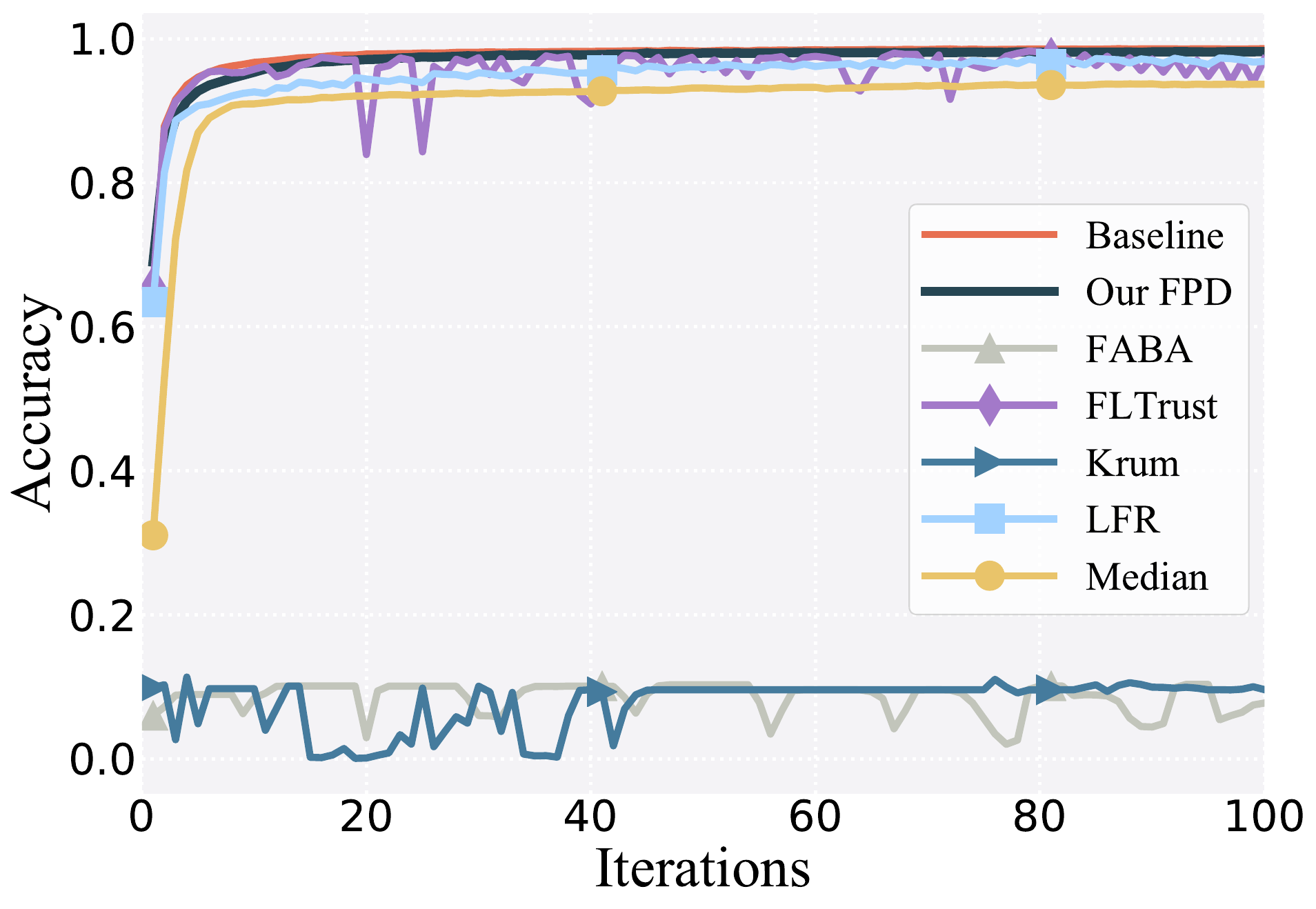}}
	\vspace{-2mm}
	\caption{Model accuracy under IPM attack}
	\label{fig:IPM}
\end{figure*}

\section{Experiments}
\subsection{Experimental Setup}
\textbf{Datasets, models, and codes.}
Our experiments are conducted on three benchmark image classification datasets: MNIST~\cite{MNIST}, Fashion-MNIST~\cite{Fashion-MNIST}, and CIFAR-10~\cite{CIFAR10}, \black{ as most of existing works did~\cite{MABRFL,momentum,Contra}}. The model structures are consistent with those in \cite{MABRFL}. \black{Our codes are available at https://github.com/CGCL-codes/FPD.}

\noindent\textbf{Data distribution.}
We follow existing works~\cite{FLTrust,FedInv} to simulate non-IID data distribution. Roughly, the non-IID degree $q\in [0,1]$ is related to the proportion of the training data with a single specific label $l\in [L]$ ($L$ is the total kinds of the labels). A larger $q$ indicates a higher non-IID degree, and $q=\frac{1}{L}$ corresponds to the IID case. In our experiments, we set $q=0.5$ by default, which is the highest non-IID setting existing works considered. Moreover, the training set sizes vary among clients. For MNIST and Fashion-MNIST, they are evenly sampled from $[10,500]$. For CIFAR-10, they are randomly chosen from $[1000,1500]$.

\noindent\textbf{Evaluated attacks.}
We consider two colluding attacks, \ie, \textit{little is enough} (LIE) attack~\cite{ALittleIsEnough}, and \textit{inner product manipulation} (IPM) attack~\cite{IPM}, as well as two non-colluding attacks, \ie, \textit{label flipping} (LF) attack~\cite{PCA}, and \textit{sign flipping} (SF) attack~\cite{WeightAttack}. \black{ Note that our defense is not limited to these attacks.} 






It is noteworthy that all the parameter settings strictly follow the recommendations stated in the original papers, as it ensures the optimal attack effectiveness.

\noindent\textbf{Evaluated defenses.}
We compare FPD with five state-of-the-art defenses, ie, Krum~\cite{Krum}, FABA~\cite{FABA}, Median~\cite{TrimmedMean}, FLTrust~\cite{FLTrust}, and LFR~\cite{LocalModelPoisoning}. Besides, we also implement FedAvg~\cite{FedAvg} in non-adversarial case as a comparison (\ie, Baseline). 


    
    

It is worth noting that these defenses rely on additional assumptions, which enhance their defense effectiveness. For example, Krum, FABA, and LFR require prior knowledge of the number of attackers to determine the number of updates to be discarded, while FLTrust and LFR depend on a clean dataset to assess the trustworthiness of updates. In contrast, our proposed FPD does not introduce any unrealistic assumptions, making it a more desirable defense for deployment in realistic scenarios with limited knowledge (\eg, just local updates).

\noindent\textbf{Performance metric and parameter settings.}
We use \textit{accuracy} (\ie, the ratio of correctly predicted samples over all the testing samples) to evaluate the performance of each defense. For a fair comparison, all the experimental results are based on the mean of three repeated experiments. We set the number of total clients $K=50$. The number of compromised clients $f=15$ by default. Each client performs $E=3$ epochs of local training for faster convergence. The prior parameters $\alpha=\beta=1$. The total iterations $T=100$. The tolerable cosine similarity $\gamma_{t}=0.8$. The importance of historical information $\lambda=0.1$. For MNIST and Fashion-MNIST, the acceptable difference between clusters $\delta=-0.1$. For CIFAR-10, $\delta=0$.

\begin{figure*}[t]
	\centering
	\subfigure[CIFAR-10]{\label{fig:lf_m}
		\includegraphics[width=0.6\columnwidth]{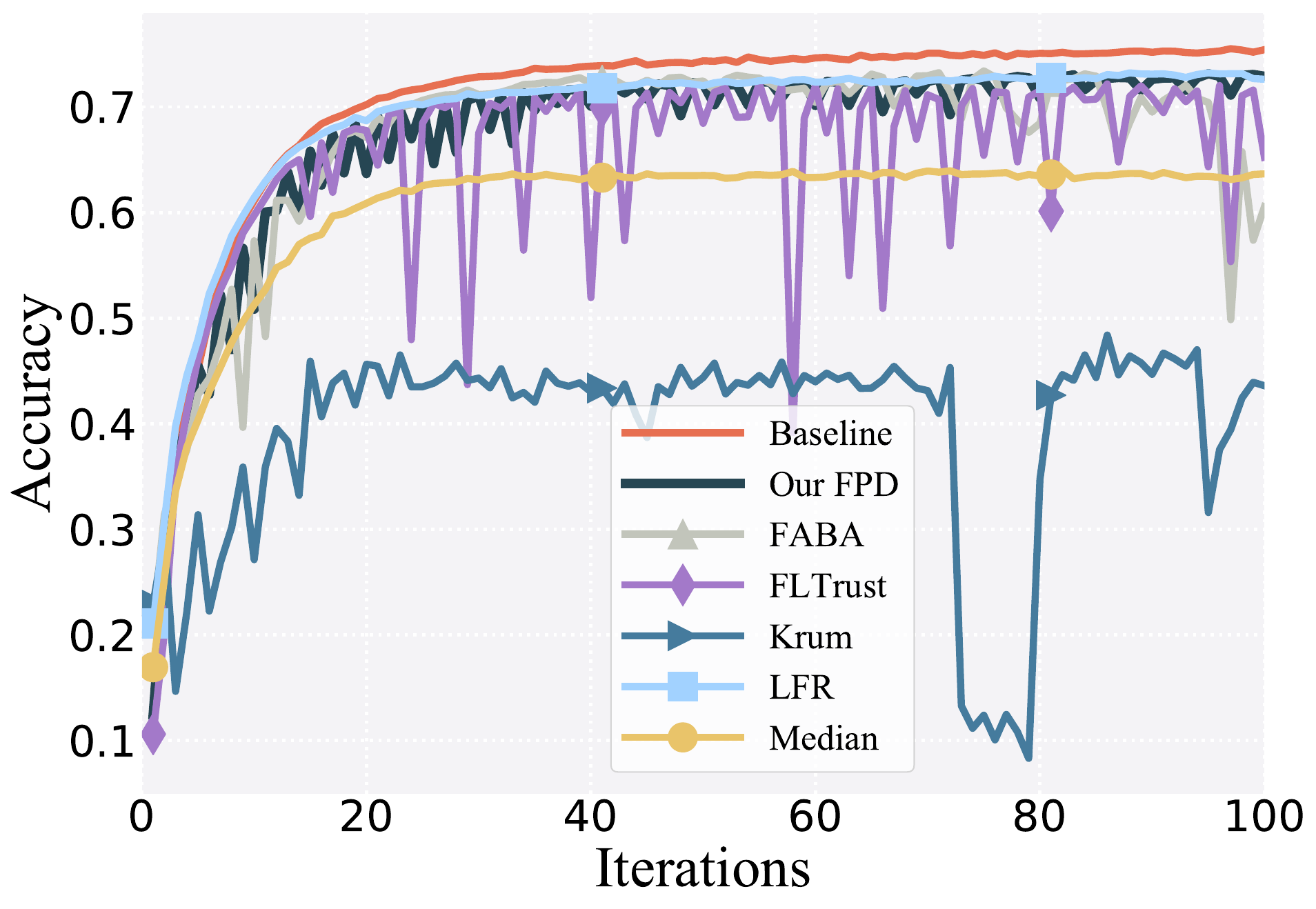}}\hspace{8mm}
	\subfigure[Fashion-MNIST]{\label{fig:lf_c}
		\includegraphics[width=0.6\columnwidth]{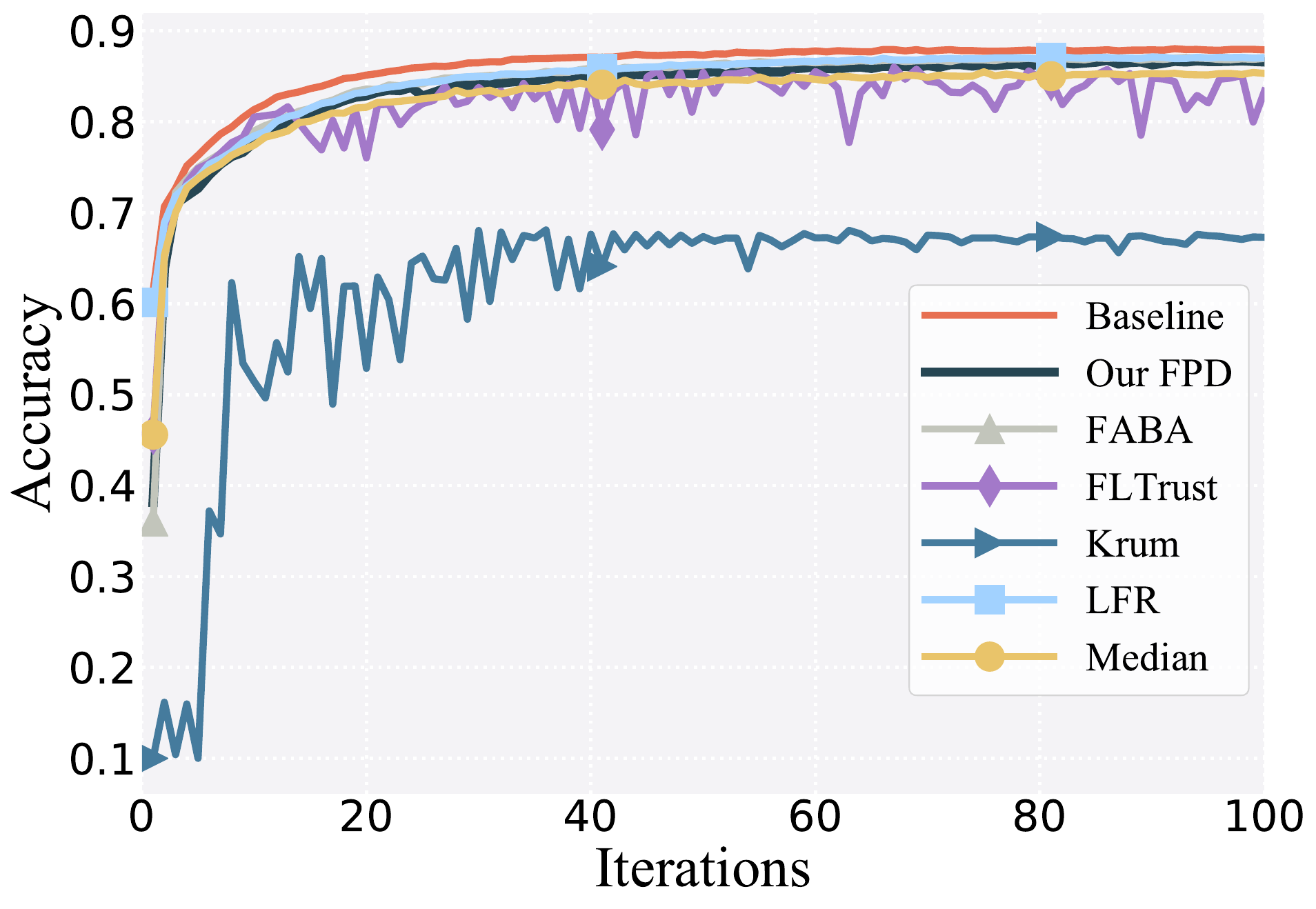}}\hspace{8mm}
	\subfigure[MNIST]{\label{fig:lf_c}
		\includegraphics[width=0.6\columnwidth]{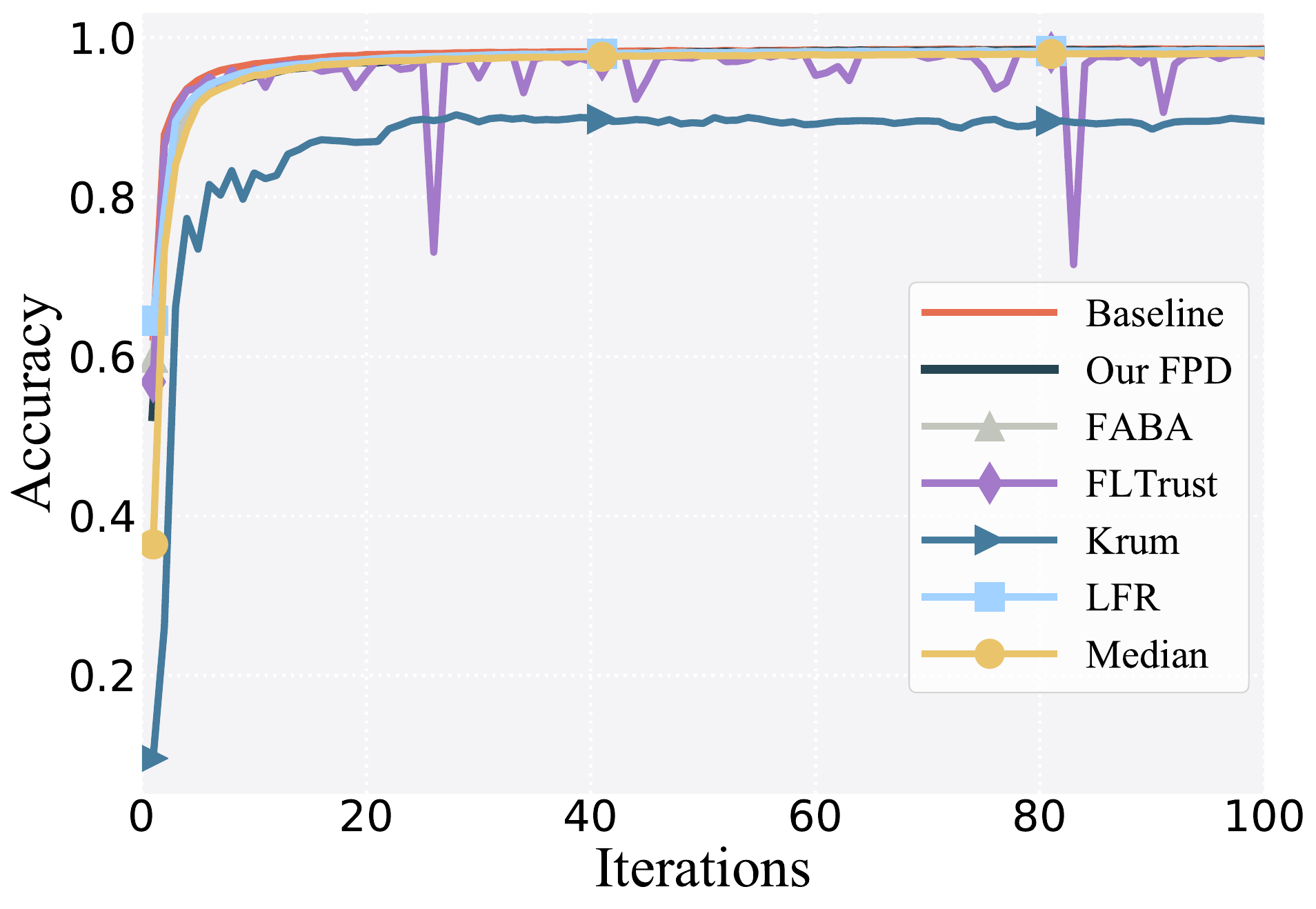}}
	\vspace{-2mm}
	\caption{Model accuracy under LF attack}
	\label{fig:LF}
\end{figure*}

\begin{figure*}[t]
	\centering
	\subfigure[CIFAR-10]{\label{fig:lf_m}
		\includegraphics[width=0.6\columnwidth]{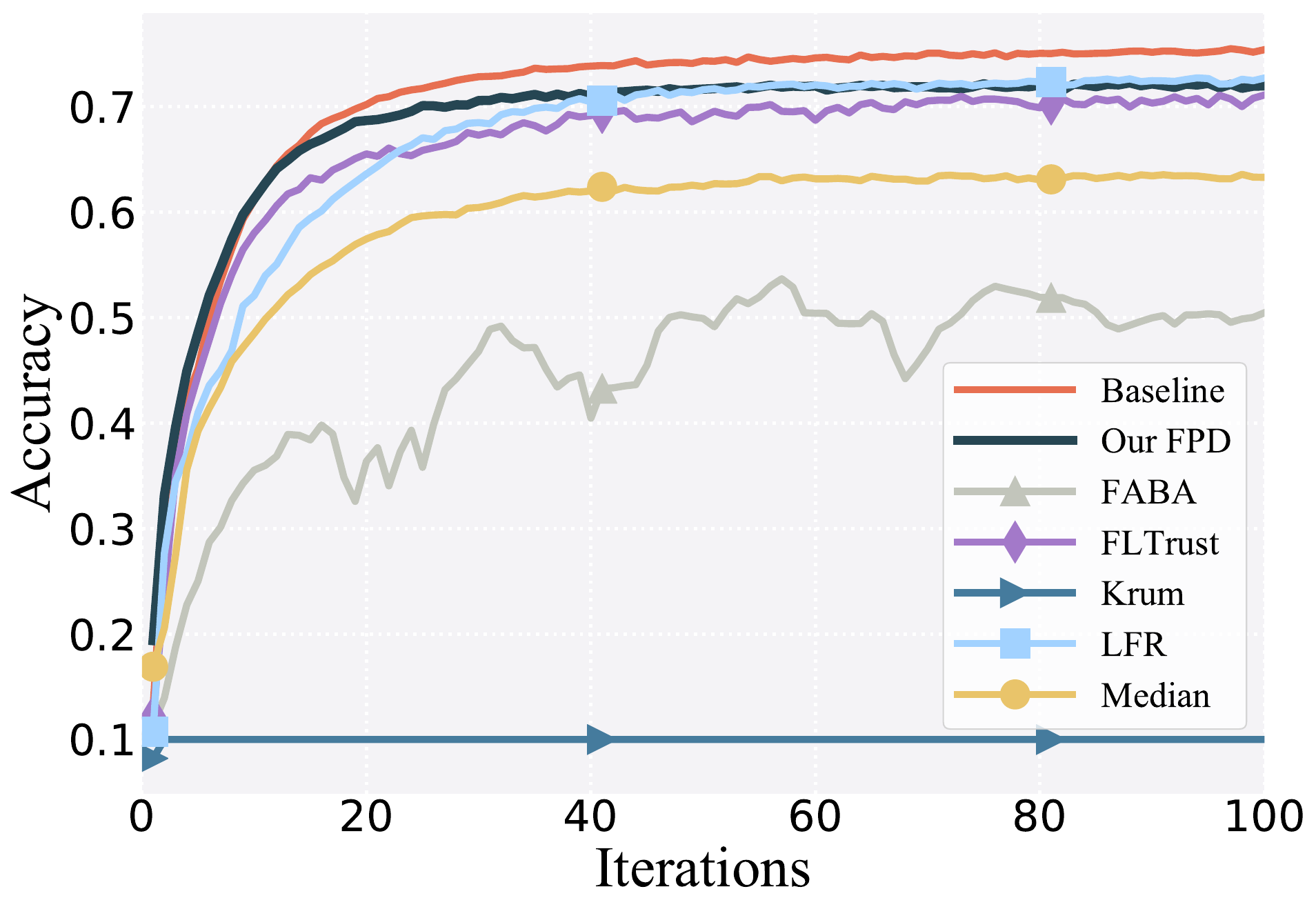}}\hspace{4mm}
	\subfigure[Fashion-MNIST]{\label{fig:lf_c}
		\includegraphics[width=0.6\columnwidth]{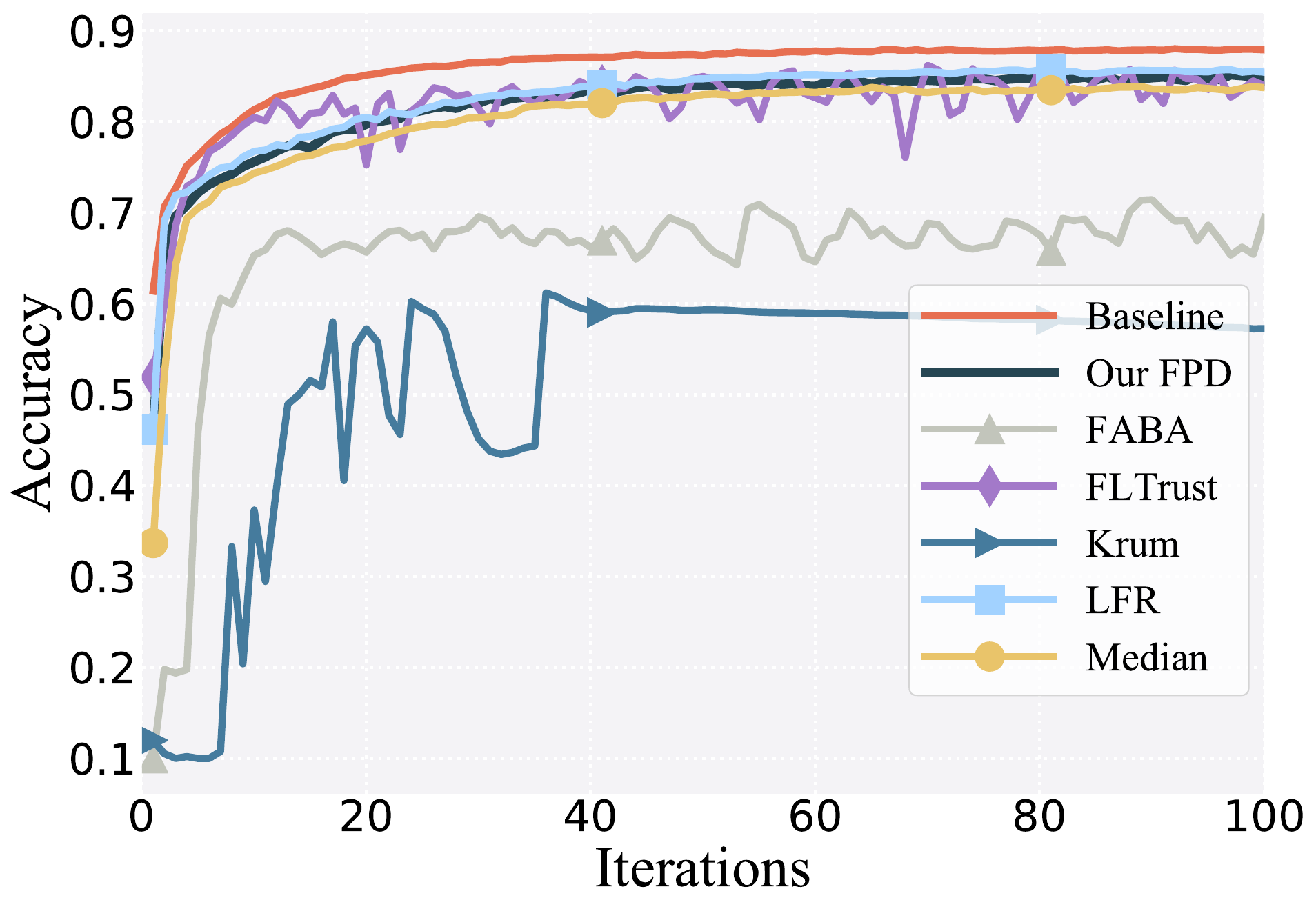}}\hspace{4mm}
	\subfigure[MNIST]{\label{fig:lf_c}
		\includegraphics[width=0.6\columnwidth]{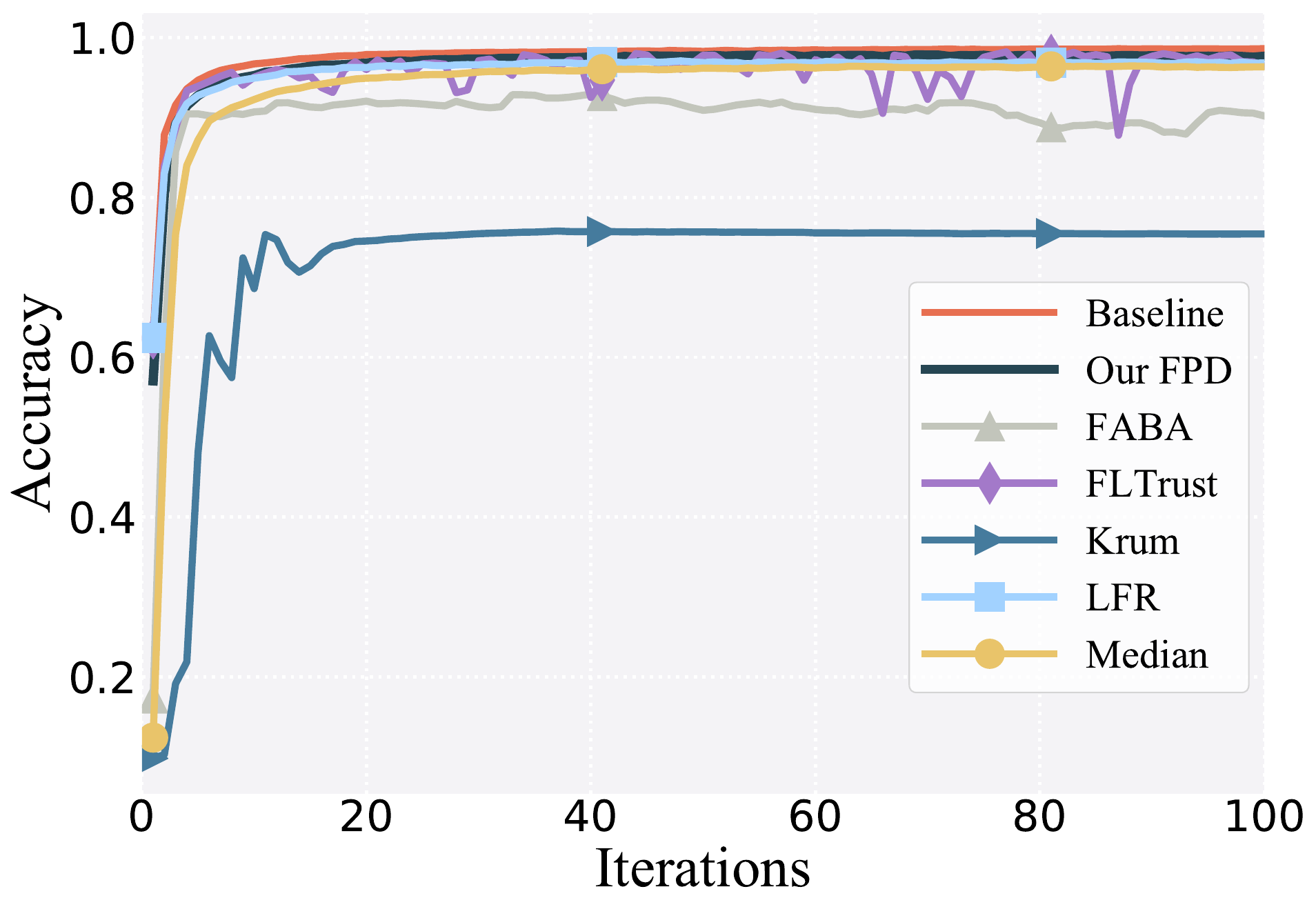}}
	\vspace{-2mm}
	\caption{Model accuracy under SF attack}
	\label{fig:SF}
\end{figure*}

\subsection{Experimental Results}

\textbf{Defense against LIE attack.}
In Fig.~\ref{fig:LIE}, we give the accuracy curves of the defenses under LIE attack on three different datasets. It is clear that the results vary across datasets. Specifically, on MNIST, Krum fail to defense. FPD, FLTrust, LFR, and Median achieve the similar accuracy with the Baseline. FABA performs slightly worse than the four defenses, with the accuracy gap of about $4\%$. On Fashion-MNIST, FPD and LFR perform best and are slightly superior to FLTrust, FABA, and Median. Krum still provides no protection. On the more complicated dataset CIFAR-10, the only defense that can effectively resist LIE attack is FPD. The other five defenses perform significantly worse than the Baseline with an accuracy gap of $20\% \sim 65\%$.

\noindent\textbf{Defense against IPM attack.}
As shown in Fig.~\ref{fig:IPM}, under IPM attack, FPD outperforms all the competitors on the three datasets with a minor gap to Baseline. Specifically, FABA and Krum are uncompetitive, because their accuracies hover at $10\%$ in all scenarios. FLTrust and LFR, which perform as well as FPD on Fashion-MNIST and MNIST, cannot defend against IPM attack on CIFAR-10. To be specific, FLTrust fluctuates sharply, and LFR converges slowly. Although Median performs much better than FABA and Krum, its accuracy is not satisfactory, especially on CIFAR-10 and Fashion-MNIST.

\noindent\textbf{Defense against LF attack.}
Fig.~\ref{fig:LF} presents the impact of LF attack on the defenses. In general, this attack is not as strong as the foregoing attacks (\ie, LIE attack and IPM attack). Specifically, FPD and LFR can perfectly shield against the attack. FLTrust and FABA can also achieve similar performance in terms of accuracy, however, they are not steady. For example, the accuracy curves of FLTrust fluctuate on all datasets (noticeably on CIFAR-10 and MNIST), and FABA suffers a drop in accuracy on CIFAR-10. Krum provides quite limited protection with lowest accuracy.

\noindent\textbf{Defense against SF attack.}
Fig.~\ref{fig:SF} shows the accuracy of the defenses under SF attack. We observe that FPD and LFR achieve the same global model accuracy, which comes near to Baseline. FLTrust is slightly inferior to the above two and incurs some fluctuation in accuracy. Median performs well on Fashion-MNIST and MNIST, however, its accuracy is about $10\%$ lower than that of FPD and LFR on CIFAR-10. FABA can partially defend against SF attack on the most fundamental MNIST dataset, nevertheless, it performs badly on CIFAR-10 and Fashion-MNIST. Krum performs worst all the time. Worse still, its accuracy on CIFAR-10 is $10\%$, which means that Krum is dispensable.

\noindent\textbf{Impact of the percentage of compromised clients.}
Table~\ref{table:impactofattackers} shows the impact of the percentage of compromised clients under LIE attack on CIFAR-10 with the non-IID degree $q=0.5$. We observe that as the percentage of attackers increases, the accuracy of all the defenses decreases. However, the degree of decreased accuracy varies from different defenses. Krum performs the worst. When there are $10\%$ attackers, its accuracy is only $43.13\%$, which is $32.38\%$ lower than the Baseline. When attackers account for $20\%$ or more, Krum fails to converge (with the accuracy of $10\%$). FABA, Median, and LFR perform better than Krum, the accuracy gap between them and the Baseline is no more than $8\%$ in the case of $10\%$ attackers. However, the gap widens significantly as the number of attackers increases to $30\%$. When attackers account for more than $30\%$, the three defenses fail to converge. FLTrust outperforms the above four defenses. When there are no more than $20\%$ attackers, FLTrust is not heavily affected, with the accuracy of about $8\%$ lower than the Baseline. We also notice that FLTrust possesses the accuracy of $46.42\%$ even in the case of $48\%$ attackers, which is drastically higher (\ie, $36.42\%$) than that of the above four defenses. However, it is about $30\%$ lower than that of the Baseline, which means that FLTrust fails to offer a satisfactory global model in high-percentage attackers scenarios. In contrast, the proposed FPD achieves the best performance all the time. More importantly, it is highly stable. Specifically, its accuracy drops from $74.81\%$ to $71.61\%$ as the fraction of attackers increases from $10\%$ to $48\%$.

\begin{table}[!t]
\renewcommand\arraystretch{0.9}
\centering
\caption{Impact of the percentage of compromised clients}
\resizebox{\linewidth}{!}{
\begin{tabular}{|c||ccccccc|}
\hline
\multirow{2}{*}{\textbf{Attackers}} & \multicolumn{7}{c|}{\textbf{Accuracy (\%)}} \\ \cline{2-8} 
 & \multicolumn{1}{c|}{\textbf{Krum}} & \multicolumn{1}{c|}{\textbf{FABA}} & \multicolumn{1}{c|}{\textbf{Median}} & \multicolumn{1}{c|}{\textbf{FLTrust}} & \multicolumn{1}{c|}{\textbf{LFR}} & \multicolumn{1}{c|}{\textbf{FPD}} & \textbf{Baseline} \\ \hline\hline
10\% & \multicolumn{1}{c|}{43.13} & \multicolumn{1}{c|}{70.43} & \multicolumn{1}{c|}{67.78} & \multicolumn{1}{c|}{71.97} & \multicolumn{1}{c|}{72.69} & \multicolumn{1}{c|}{\textbf{74.81}} & \multirow{6}{*}{75.51} \\ \cline{1-7}
20\% & \multicolumn{1}{c|}{10.00} & \multicolumn{1}{c|}{63.44} & \multicolumn{1}{c|}{59.68} & \multicolumn{1}{c|}{68.07} & \multicolumn{1}{c|}{57.60} & \multicolumn{1}{c|}{\textbf{74.54}} &  \\ \cline{1-7}
30\% & \multicolumn{1}{c|}{10.00} & \multicolumn{1}{c|}{36.50} & \multicolumn{1}{c|}{50.47} & \multicolumn{1}{c|}{56.20} & \multicolumn{1}{c|}{45.01} & \multicolumn{1}{c|}{\textbf{73.43}} &  \\ \cline{1-7}
40\% & \multicolumn{1}{c|}{10.00} & \multicolumn{1}{c|}{10.00} & \multicolumn{1}{c|}{10.00} & \multicolumn{1}{c|}{48.73} & \multicolumn{1}{c|}{10.00} & \multicolumn{1}{c|}{\textbf{72.51}} &  \\ \cline{1-7}
44\% & \multicolumn{1}{c|}{10.00} & \multicolumn{1}{c|}{10.00} & \multicolumn{1}{c|}{10.00} & \multicolumn{1}{c|}{48.06} & \multicolumn{1}{c|}{10.00} & \multicolumn{1}{c|}{\textbf{72.02}} &  \\ \cline{1-7}
48\% & \multicolumn{1}{c|}{10.00} & \multicolumn{1}{c|}{10.00} & \multicolumn{1}{c|}{10.00} & \multicolumn{1}{c|}{46.42} & \multicolumn{1}{c|}{10.00} & \multicolumn{1}{c|}{\textbf{71.61}} &  \\ \hline
\end{tabular}
}
\label{table:impactofattackers}
\end{table}

\begin{table}[!t]
\renewcommand\arraystretch{0.9}
\centering
\caption{Impact of the non-IID degree}
\resizebox{\linewidth}{!}{
\begin{tabular}{|c||ccccccc|}
\hline
\multirow{2}{*}{\textbf{\begin{tabular}[c]{@{}c@{}}Non-IID\\  Degree\end{tabular}}} & \multicolumn{7}{c|}{\textbf{Accuracy (\%)}} \\ \cline{2-8} 
 & \multicolumn{1}{c|}{\textbf{Krum}} & \multicolumn{1}{c|}{\textbf{FABA}} & \multicolumn{1}{c|}{\textbf{Median}} & \multicolumn{1}{c|}{\textbf{FLTrust}} & \multicolumn{1}{c|}{\textbf{LFR}} & \multicolumn{1}{c|}{\textbf{FPD}} & \textbf{Baseline} \\ \hline\hline
0.1 & \multicolumn{1}{c|}{10.00} & \multicolumn{1}{c|}{48.48} & \multicolumn{1}{c|}{56.93} & \multicolumn{1}{c|}{71.88} & \multicolumn{1}{c|}{\textbf{75.52}} & \multicolumn{1}{c|}{75.15} & 76.91 \\ \hline
0.3 & \multicolumn{1}{c|}{10.00} & \multicolumn{1}{c|}{47.29} & \multicolumn{1}{c|}{55.45} & \multicolumn{1}{c|}{71.33} & \multicolumn{1}{c|}{\textbf{75.51}} & \multicolumn{1}{c|}{75.00} & 75.89 \\ \hline
0.5 & \multicolumn{1}{c|}{10.00} & \multicolumn{1}{c|}{36.50} & \multicolumn{1}{c|}{50.47} & \multicolumn{1}{c|}{56.20} & \multicolumn{1}{c|}{45.01} & \multicolumn{1}{c|}{\textbf{73.43}} & 75.51 \\ \hline
0.7 & \multicolumn{1}{c|}{10.00} & \multicolumn{1}{c|}{10.00} & \multicolumn{1}{c|}{10.00} & \multicolumn{1}{c|}{47.49} & \multicolumn{1}{c|}{33.54} & \multicolumn{1}{c|}{\textbf{71.55}} & 71.85 \\ \hline
0.9 & \multicolumn{1}{c|}{10.00} & \multicolumn{1}{c|}{10.00} & \multicolumn{1}{c|}{10.00} & \multicolumn{1}{c|}{28.31} & \multicolumn{1}{c|}{10.00} & \multicolumn{1}{c|}{\textbf{60.31}} & 61.79 \\ \hline
0.95 & \multicolumn{1}{c|}{10.00} & \multicolumn{1}{c|}{10.00} & \multicolumn{1}{c|}{10.00} & \multicolumn{1}{c|}{23.58} & \multicolumn{1}{c|}{10.00} & \multicolumn{1}{c|}{\textbf{53.61}} & 54.41 \\ \hline
\end{tabular}
}
\label{table:impactofnoniid}
\end{table}

\noindent\textbf{Impact of the non-IID degree.}
Table~\ref{table:impactofnoniid} shows the impact of the non-IID degree under LIE attack on CIFAR-10 with $30\%$ compromised clients. We observe that as the non-IID degree $q$ varies from 0.1 (\ie, the IID case) to 0.95 (\ie, the extremely non-IID case), all the schemes (including Baseline) achieve a lower and lower accuracy gradually. However, the accuracy of FPD is invariably comparable with that of Baseline (with the accuracy gap of $0.30\% \sim 2.08\%$). FLTrust and LFR perform well when $q=0.1$ and $q=0.3$. However, when $q \geq 0.5$, their accuracy drops dramatically, which indicates that FLTrust and LFR do not apply to non-IID scenario. Krum, FABA, and Median cannot obtain a high-quality global model even in IID setting (\ie, $q=0.1$) due to the remarkable attack effect of LIE attack.


\noindent\textbf{Ablation study on the absence of modules.}
We perform an ablation study to understand the empirical effects of different modules in Table~\ref{table:ablationstudy}, where $A, B, C, D$ indicate \textit{reliable client selection, mitigating colluding attacks, mitigating non-colluding attacks,} and \textit{update denoising} respectively. It can be seen that without module $A$ the global model accuracy decreases $0.57\%\sim3.37\%$, and without module $D$ the global model accuracy decreases $1.39\%\sim3.01\%$, which indicates that the two modules can slightly improve off-the-shelf defenses. Without module $B$, the global model accuracy under LIE attack drops $4.85\%$, which means that module $B$ is effective to defend against colluding attacks. Without module $C$, the combination cannot achieve a desirable global model accuracy under non-colluding attacks (\ie, LF and SF attacks), demonstrating the necessity of module $C$.

\begin{table}[!t]
\tiny
\renewcommand\arraystretch{1}
\centering
\caption{Ablation study on CIFAR-10 with $30\%$ attackers}
\vspace{-0mm}
\resizebox{1\linewidth}{!}{
\begin{tabular}{|c||cccc|}
\hline
\multirow{2}{*}{\textbf{Combination}} & \multicolumn{4}{c|}{\textbf{Accuracy (\%)}}                                                           \\ \cline{2-5} 
                             & \multicolumn{1}{c|}{\textbf{LIE}}   & \multicolumn{1}{c|}{\textbf{IPM}}   & \multicolumn{1}{c|}{\textbf{LF}}    & \textbf{SF}    \\ \hline \hline
A+B+C+D                      & \multicolumn{1}{c|}{73.43} & \multicolumn{1}{c|}{73.96} & \multicolumn{1}{c|}{74.26} & 73.42 \\ \hline
B+C+D                        & \multicolumn{1}{c|}{72.86} & \multicolumn{1}{c|}{72.52} & \multicolumn{1}{c|}{72.38} & 70.05 \\ \hline
A+C+D                        & \multicolumn{1}{c|}{68.58} & \multicolumn{1}{c|}{71.71} & \multicolumn{1}{c|}{73.89} & 73.34 \\ \hline
A+B+D                        & \multicolumn{1}{c|}{72.30} & \multicolumn{1}{c|}{72.43} & \multicolumn{1}{c|}{63.76} & 67.96 \\ \hline
A+B+C                        & \multicolumn{1}{c|}{71.42} & \multicolumn{1}{c|}{72.57} & \multicolumn{1}{c|}{71.89} & 71.41 \\ \hline
\end{tabular}
}
\label{table:ablationstudy}
\end{table}

\noindent\textbf{Performance under mixed attack.}
Previous experiments have demonstrated that FPD exhibits superior defense performance against individual colluding attacks or non-colluding attacks. As a result, one may naturely wonder whether FPD can withstand \textit{mixed attacks} (MA) as well, \ie, a group of attackers deploy colluding attacks while the remaining deploy non-colluding attacks. To this end, we conduct MA (half of attackers deploy LIE and the other half deploy LF) and compare it with LIE and LF, the results are shown in Tab.~\ref{table:mixed attack}. Surprisingly, MA is not stronger than LIE, and sometimes even weaker than LF. Specifically, our FPD performs consistently well under the three attacks with the highest accuracy, demonstrating its superiority in eliminating malicious updates. For FLTrust and LFR, MA is somewhat effective, but its impact is intermediate between that of pure LIE and LF. This is because both defenses are effective in defending against LF, but are weak in identifying LIE attackers. As for Krum, FABA, and Median, MA has the slightest effect on accuracy, we speculate that MA makes malicious updates more dispersed, thus making it easier for these similarity-based defenses to identify benign updates.

\begin{table}[!t]
\renewcommand\arraystretch{0.9}
\centering
\caption{Performance under MA on CIFAR-10 with $30\%$ attackers}
\resizebox{\linewidth}{!}{
\begin{tabular}{|c||cccccc|}
\hline
\multirow{2}{*}{\textbf{Attack}} & \multicolumn{6}{c|}{\textbf{Accuracy (\%)}}                                                                                                                                                               \\ \cline{2-7} 
                                 & \multicolumn{1}{c|}{\textbf{Krum}} & \multicolumn{1}{c|}{\textbf{FABA}} & \multicolumn{1}{c|}{\textbf{Median}} & \multicolumn{1}{c|}{\textbf{FLTrust}} & \multicolumn{1}{c|}{\textbf{LFR}} & \textbf{FPD} \\ \hline\hline
LIE                              & \multicolumn{1}{c|}{10.00}         & \multicolumn{1}{c|}{36.50}         & \multicolumn{1}{c|}{50.47}           & \multicolumn{1}{c|}{56.20}            & \multicolumn{1}{c|}{45.10}        & 73.43        \\ \hline
LF                               & \multicolumn{1}{c|}{43.60}         & \multicolumn{1}{c|}{60.55}         & \multicolumn{1}{c|}{63.66}           & \multicolumn{1}{c|}{69.29}            & \multicolumn{1}{c|}{72.62}        & 73.14        \\ \hline
MA                               & \multicolumn{1}{c|}{48.97}           & \multicolumn{1}{c|}{65.32}         & \multicolumn{1}{c|}{64.82}           & \multicolumn{1}{c|}{62.44}              & \multicolumn{1}{c|}{59.13}        & 73.46          \\ \hline
\end{tabular}
}
\label{table:mixed attack}
\end{table}

\section{Limitations}
Although our proposed FPD performs best,  there are still some limitations.

\noindent\textbf{Suboptimal performance when attackers dominate.}
Our defense suffers from an accuracy degrade when attackers dominate. Because the server lacks a gold standard, the server can only assume that the majority is reliable as did in existing defenses~\cite{FABA,TrimmedMean,MABRFL}. Though some works (\eg, FLTrust) work in such an extreme case, they make a stronger assumption, \ie, the server owns a clean dataset, which obviously violates the privacy requirements of FL.

\noindent\textbf{Lack of theoretical analysis.}
In the literature of security studies in federated learning, it is difficult to provide a theoretical security analysis~\cite{FedInv,MABRFL}, and our scheme is also heuristic. It is a  challenging and promising topic and we leave it to our future work.

\section{Conclusion}
This paper proposed FPD, a four-pronged defense against Byzantine attacks. Specifically, FPD first performs reliable client selection to encourage participants to share high-quality updates. Next, a similarity-based filter is employed to prohibit the adversary from designing excessively similar malicious updates, enhancing the difficulty of launching a covert attack. Then, FPD utilizes a spectral-based outlier detector to remove the updates far from the overall distribution. Finally, an autoencoder is used to denoise the slightly noisy but harmful updates. Extensive experiments demonstrate that FPD is superior to existing defenses.

\section*{Acknowledgments} Shengshan's work is supported in part by the National Natural Science Foundation of China (Grant No.U20A20177) and Hubei Province Key R\&D Technology Special Innovation Project under Grant No.2021BAA032. 
Minghui's work is supported in part by the National Natural Science Foundation of China (Grant No. 62202186)
Shengshan Hu is the corresponding author.

\bibliographystyle{ACM-Reference-Format}
\balance
\footnotesize
\bibliography{acmart}


\end{document}